\documentclass[letter,openany]{aa}   
\usepackage{natbib}
\bibpunct{(}{)}{;}{a}{}{,} 
\usepackage{graphicx}
\usepackage{xcolor}
\usepackage[colorlinks,allcolors=blue]{hyperref}

\usepackage{txfonts}
\usepackage{orcidlink}
\usepackage{ulem}

\begin{document}

   \title{First detection of the $J_{-1}\rightarrow(J-$ 1)$_{0}-E$ methanol maser transitions at $J=7$ and $10$}

   \author{Pedro K. Humire\orcidlink{0000-0003-3537-4849} \inst{\ref{inst.USP},\ref{inst.MPIfR},\ref{inst.JAO},\ref{inst.USM}}
   \and Gisela N.\ Ortiz-León\orcidlink{0000-0002-2863-676X} \inst{\ref{inst.INAUE},\ref{inst.MPIfR}}
  \and Antonio Hernández-Gómez\orcidlink{0000-0001-7520-4305} \inst{\ref{inst.TECH}}
   \and Wenjin Yang\orcidlink{0000-0002-3599-6608} \inst{\ref{inst.Nanjing},\ref{inst.MPIfR}}
   \and Christian Henkel\orcidlink{0000-0002-7495-4005} \inst{\ref{inst.MPIfR},\ref{inst.Xinjiang}}
   \and Sergio Martín\orcidlink{0000-0001-9281-2919} \inst{\ref{inst.ESOChile}}
    }

   \institute{\label{inst.USP}Departamento de Astronomia, Instituto de Astronomia, Geofísica e Ciências Atmosféricas da USP, Cidade Universitária, 05508-900 São Paulo, SP, Brazil,
   \label{email}pedrokhumire@usp.br
   \and \label{inst.MPIfR}Max-Planck-Institut f\"ur Radioastronomie, Auf-dem-H\"ugel 69, 53121 Bonn, Germany,
   \and \label{inst.JAO}Joint ALMA Observatory, Alonso de Cordova 3107, Vitacura, Santiago, Chile, 
   \and \label{inst.USM}Departamento de Física, Universidad Técnica Federico Santa María, Casilla 110-V, Avda. España 1680, Valparaíso, Chile
   \and \label{inst.INAUE}Instituto Nacional de Astrofísica, Óptica y Electrónica, Apartado Postal 51 y 216, 72000 Puebla, México
   \and \label{inst.TECH}Tecnologico de Monterrey, Escuela de Ingeniería y Ciencias, Avenida Eugenio Garza Sada 2501, Monterrey 64849, Mexico
   \and \label{inst.Nanjing}School of Astronomy \& Space Science, Nanjing University, 163 Xianlin Avenue, Nanjing 210023, People's Republic of China
    \and\label{inst.Xinjiang}Xinjiang Astronomical Observatory, Chinese Academy of Sciences, 830011 Urumqi, China
   \and \label{inst.ESOChile}European Southern Observatory, Alonso de C\'ordova, 3107, Vitacura, Santiago 763-0355, Chile  
   }

   \date{Received March 20, 2024; accepted March 25, 2024}

 
  \abstract
   {Class\,I methanol masers provide sensitive information about the shocked environment around star-forming regions. Among the brightest Class~I methanol masers, we have those in the $J_{-1}\rightarrow(J-$ 1)$_{0}-E$ line series, currently reported for the $J=4-9$ transitions, with the only exception being the $J=7$ one at 181.295~GHz, and never expanded to higher $J$ transitions.} 
   {We aim to search for population inversion in the $7_{-1}\rightarrow6_{0}-E$ and $10_{-1}\rightarrow9_{0}-E$ methanol transition lines at 181.295 and 326.961\,GHz, respectively, and also extend the number of known low-mass star-forming sources harboring Class\,I methanol masers.}
   {We employed the 
   Atacama Pathfinder Experiment (APEX) 12\,m telescope to survey low-mass Galactic sources, focusing on methanol emission lines. We conducted rotation diagrams for all sources with detected $J=7$ methanol line transitions, while employing radiative transfer modeling (both in and out of local thermodynamic equilibrium) to characterize methanol excitation conditions in detail for one specific source with detected masers.
   }
   {We have detected the $7_{-1}\rightarrow6_{0}-E$ and $10_{-1}\rightarrow9_{0}-E$ methanol transitions in six out of nineteen sources. Among them, we firmly determined the $10_{-1}\rightarrow9_{0}-E$ maser nature in CARMA\,7, L1641N, NGC\,2024, and Serpens FIRS, and we claim for the presence of inverted population emission in the $7_{-1}\rightarrow6_{0}-E$ line toward CARMA\,7 and L1641N. This represents the first report of methanol maser emission in these particular transitions. Our study supports previous works indicating that conditions for Class\,I methanol maser emission are satisfied in low-mass star-forming regions and expands the range of detectable frequencies toward higher values.}
   {}

   \keywords{stars: formation -- stars: low-mass -- stars: protostars -- ISM: molecules -- ISM: masers}

   \maketitle
%

\section{Introduction}
\label{sec.intro}
Methanol (CH$_{3}$OH) emission is quite common in the radio regime, with hundreds of transitions in the sub-mm window \citep[e.g.,][]{Comito2005}. This molecule is also prone to population inversion under specific excitation conditions, causing maser emission \citep[e.g.,][]{Cragg1992}. In particular, methanol masers (MM) constitute unique tools for studying the physical properties of dense gas associated with young stellar objects (YSOs). Given their brightness and compactness \citep[e.g.,][]{Menten1991b}, their positions can be determined with high-precision astrometry and over vast distances \citep[e.g.,][]{Xu2021}.

It was early discovered that MM can be divided into two classes, a collisionally pumped Class\,I (MMcIs) and a radiatively pumped Class\,II (MMcIIs) \citep{Batrla1987,Menten1991a}, which trace shocked regions, such as outflows (MMcIs) or the circumstellar environment close to central massive young stellar objects (MMcIIs), such as their protoplanetary disks. MMcIs have been detected toward high and low-mass stars \citep{Kalenskii2006,Kalenskii2010,Kalenskii2013,Kalenskii2017,Rodriguez2017,Yang2023}, while MMcIIs have been detected only toward high-mass YSOs \citep[e.g.,][]{Minier2003,Breen2013}. MMcIs are the only class definitely detected in emission beyond the Local Group of galaxies (i.e., beyond 1.5~Mpc) \citep{Gorski2018,Humire2020,McCarthy2021,Humire2022}, while absorption features have been reported for the Seyfert~2 system NGC~3079 \citep{Impellizzeri2008}. There is also a tentative detection of MMcIIs at these distances \citep{Chen2022}.

Methanol maser emission in the $J_{-1}\rightarrow(J-$ 1)$_{0}-E$ series (hereafter MMcIs$J_{-1}$) has been only detected in the $J=4-9$ transition range because low-$J$ ($J<$4) transitions are anti-inverted and higher $J$ transitions have never been encountered as masers. The difficulty in finding inverted populations in high-$J$ transitions lies in the correlation between frequency ($\nu$) and the time required to accumulate the inverted population ($\nu^{-3}\propto$ A$^{-1}$, where A is the Einstein coefficient for spontaneous emission), making population inversion progressively harder to achieve in proportion to the cube of the frequency \citep[for the case of hydrogen, see, e.g.,][their Appendix\,A]{Scoville2013}.

Accounting for the first detections of MMcIs$J_{-1}$, the $J=4$ and $J=5$ transitions in this line series (at 36 and 84\,GHz) were first detected in space in the early 70's \citep{Turner1972,Zuckerman1972}. These transitions were later identified as masers in the 80's by \citet{Morimoto1985} and \citet{Batrla1988}, respectively. The first report of maser emission for the $J=6$ transition at 132.9\,GHz was made ten years later by \citet{Slysh1997}, followed by the $J=8$ transition at 229\,GHz by \citet{Slysh2002}. A decade ago, \citet{Yanagida2014} detected the $J=9$ transition (at 278.3\,GHz), which remains the highest $J$ transition detected as a maser in the $J_{-1}\rightarrow(J-$ 1)$_{0}-E$ line series.

Despite the long-term history of research, not much attention has been paid to the $J=7$ transition at 181.295\,GHz. The mere detection of this line would represent a unique 16.7\% (1/6) factor improvement in the number of detected maser lines for the MMcIs$J_{-1}$ series of emission lines, if we stick to the range of transitions already detected. Maser models are expected to improve their capability of reproducing observations through synthetic spectra \citep[see, e.g.,][]{Lee2023,Xue2024}, an accuracy of 16.7\% better than currently available is thus desired. 

Among the $J_{-1}\rightarrow(J-$ 1)$_{0}-E$ methanol transition series, detecting the lines at $J=$7 and $J=$10 poses significant challenges due to their proximity to the telluric water lines at 181.310 and 325.153\,GHz, respectively. This necessitates exceptionally favorable weather conditions to mitigate low atmospheric transmission and high noise levels during observations. However, successfully detecting these lines can benefit future observing campaigns targeting the entire $J_{-1}\rightarrow(J-$ 1)$_{0}-E$ maser series within the same sources.

A strong argument for searching not only for the $J=7$ transition but also for higher-$J$ transition lines relates to the interstellar medium (ISM) conditions we are tracing since specific $J$ transitions of the same molecular species emerge at a certain restricted density range \citep{Shirley2015,Leurini2016}. The advent of improved facilities and atmospheric conditions in the sub-mm regime allows us to also search for the $J=$10 transition, with an angular resolution that, for a given telescope, is almost twice as good (on a linear scale) as the one of its $J=7$ sibling.

In the following sections, we will present our observations as well as details of the data reduction and introduce our sample of sources (Sect.\,\ref{Sec.Observations}). Sect.\,\ref{Sec.Results} describes the two methods used to identify methanol masers and presents our results in the context of the latest advances in the field. Finally, we conclude in Sect.\,\ref{Sec.Conclusions}.

\section{Observations}
\label{Sec.Observations}

\subsection{The sample}
\label{subsec.subsample}

We have conducted a survey of 19 low-mass star-forming regions searching for the $7_{-1}\rightarrow 6_{0}-E$ and $10_{-1}\rightarrow 9_{0}-E$ methanol transitions. The sample consists of the following sources: CARMA-7, L1641N, NGC\,2024, IRAS\,16293, Serpens\,FIRS, Serpens\,SMM4, YLW 16A, NGC 2027 North, Orion A West, VLA 1623, HH1-2 VLA 1, HH 212, Haro 4-255, GSS 30-IRS1, L483, L483 FIR, HOPS 96, IRAS 18264-0143, and LDN 723-mm.





\subsection{SEPIA180 and SEPIA345 observations}
\label{Sec.observations}
The observations were carried out with the Atacama Pathfinder EXperiment (APEX) 12\,m telescope located in the Llano Chajnantor, Chile \citep{Guesten2006}, under projects M-0109.F-9512B-2022 and M-0109.F-9512C-2022 (P.I. A. Hern\'andez-G\'omez) at different dates: 1$^{\rm{st}}$ April, 24-27$^{\rm{th}}$ May and 17$^{\rm{th}}$ June 2022. We used the Swedish ESO PI receiver SEPIA \citep{Belistsky2018}, a double sideband (2SB) dual-polarization receiver working in a frequency range between 159-211 GHz. The backend was a Fast Fourier Transform fourth-generation spectrometer (FFTS4G) that consisted of two sidebands with 4 GHz bandwidth each, covering a total bandwidth of 8 GHz. The observations were taken using the wobbler-switching mode with a switching frequency of 1.7\,Hz and a beam throw of 120$''$.

Here we used two frequency setups for the observations. With  SEPIA180 we covered a frequency range of 181.140 to 185.140~GHz in the upper side band (USB) and 193.480 to 197.480 GHz in the lower side band (LSB). With SEPIA345 we covered a frequency range of 323.153 to 327.153~GHz in the USB and of 335.153 to 339.153~GHz in the LSB. The spectrometer has 65536 channels, providing a spectral resolution of 61\,kHz, corresponding to a velocity resolution of $\sim 0.1$~km~s$^{-1}$ at 183.310 GHz and $\sim 0.06$~km~s$^{-1}$ at 325.153 GHz. We applied the {\tt resample} task inside GILDAS/CLASS\footnote{\url{http://www.iram.fr/IRAMFR/GILDAS}} to produce a common 0.1~km~s$^{-1}$ channel width for both setups. This initial common velocity resolution changed after smoothing (see below).

The total observing time on each source varied slightly among sources depending on the precipitable water vapor (pwv) during the observations. The pwv varied between 0.023 and 1.14~mm. On average, the on-source time was $\sim$15 minutes with SEPIA180 and $\sim25$ minutes with SEPIA345. The system temperatures varied between 78 and 721~K during the observations. The following sources were used for focus and pointing calibrations: VY-CMA, IRC+10216, R-LEP, RAFGL 4211, IRAS 15194-51, IRC+00365, G327-ATCA, IRC+20370, and RAFGL 1922.

The calibrated data is delivered in antenna temperature units ($T_{\rm{A}}^*$). To convert it to main beam brightness $T_{\rm{MB}}$, we have used the relation $T_{\rm{MB}} = T_{\rm{A}}^*(\eta_{\rm{fw}}/\eta_{\rm{MB}})$, where $\eta_{\rm{fw}}$ is the forward coupling efficiency (0.95) and $\eta_{\rm{MB}}$ is the main-beam efficiency. The main beam efficiency depends on the aperture efficiency $\eta_{\rm{a}}$, which has a value of 0.71 for SEPIA180 and 0.67 for SEPIA345 based on observations toward Mars\footnote{The efficiencies can be found at \url{http://www.apex-telescope.org/telescope/efficiency/index.php}}. Since $\eta_{\rm{MB}} = 1.2182 \times \eta_{\rm{a}}$, then $\eta_{\rm{MB}} =0.865$ for SEPIA180 and $\eta_{\rm{MB}} =0.816$ for SEPIA345.

The absolute calibration uncertainty is estimated to be $\sim10$\% \citep{Dumke2010}.
\noindent The half-power beam width (HPBW) at 181.310~GHz is 34$''$, while it is 19.2\arcsec\ at 325.153~GHz.

\subsection{Data reduction}
\label{Sec.data_reduction}
The data reduction was performed with the GILDAS/CLASS software developed by the Institut de Radioastronomie Millim\'etrique (IRAM). 
A first-order baseline was applied to the spectra by selecting windows free of line emission and subtracting them from the data. The resulting spectra were later smoothed using a Box Kernel\footnote{see \url{ https://docs.astropy.org/en/stable/api/astropy.convolution.Box1DKernel.html}} of width$=$20, averaging the data over 20 points to reduce fluctuations and to enhance the signal-to-noise ratio in individual channels. Comparing the estimated radial and line-width velocities from a single Gaussian before and after smoothing, we see a difference of $\leq$0.2~km~s$^{-1}$ in central velocities and $\leq$0.6~km~s$^{-1}$ in line-widths. We, therefore, estimate a conservative new velocity resolution of 1~km~s$^{-1}$.

\begin{figure*}[!ht]
\centering
\includegraphics[width=0.99\textwidth, trim={0 0 0 0}, clip]{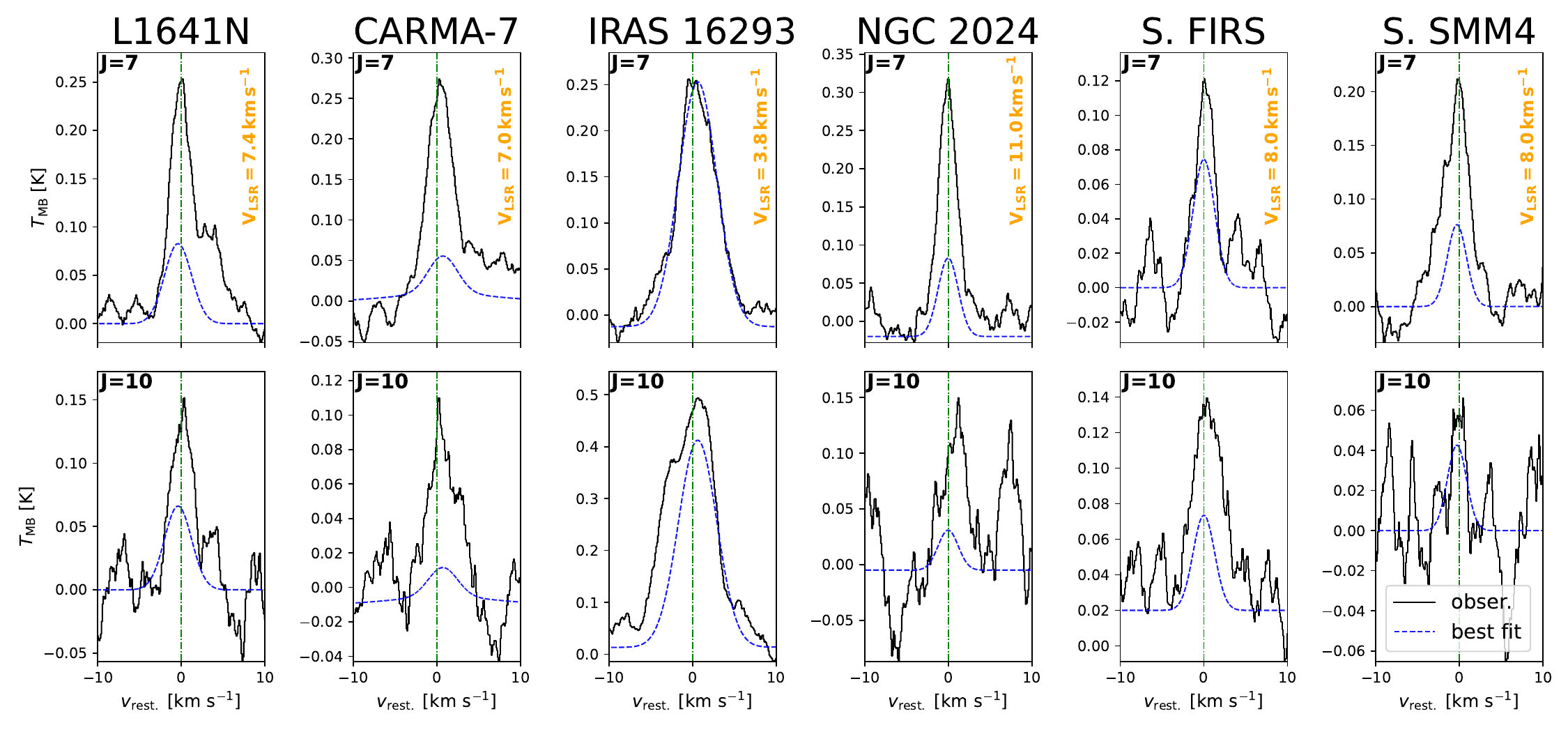}
\caption{Rest frame $J_{-1}\rightarrow(J-$ 1)$_{0}-E$ methanol $J=$7 and 10 transition spectra of our sub-sample (Sect.\,\ref{subsec.subsample}, solid black lines), along with the rotation diagrams best fits (dashed blue lines, see Fig.\,\ref{Fig.1}). The Local Standard of Rest velocities ($V_{\rm{LSR}}$; labeled in orange) of the different sources were obtained from the literature where we assume a common uncertainty of 1.0~km~s$^{-1}$ (see Sect.\,\ref{Sec.data_reduction}).}
\label{Fig.maser_specs}
\end{figure*}

During our observational proposal process, we obtained the Local Standard of Rest velocities ($V_{\rm{LSR}}$) for our entire sample from the literature, which was then recorded in the headers of our APEX files. The Local Standard of Rest velocities ($V_{{\rm{LSR}}}$) for our sub-sample are highlighted with vertical orange labels in Fig.\,\ref{Fig.maser_specs}. Specifically, we referenced \citet{Podio2021} for CARMA-7, \citet{Bae2011} for L1641N, \citet{Pineda2012} for IRAS\,16293, \citet{Buckle2010} for NGC\,2024, \citet{Plunkett2015} for Serpens~FIRS, and \citet{Narayanan2002} for Serpens~SMM4. Due to an oversight, we mistakenly assigned a velocity of 7 instead of 8.1~km~s$^{-1}$ to CARMA-7. CASSIS later detected this error, and its line fitting provided, after several modeling iterations, a corrected velocity of $\sim$7.8~km~s$^{-1}$, which should be adopted as the real $V_{{\rm{LSR}}}$ of this source.

Following Section \ref{Sec.observations}, we adopt a conservative approach by assuming a uniform velocity uncertainty of 1.0~km~s$^{-1}$ across the entire dataset. To contextualize this value, it is larger than the statistical uncertainties derived from Markov chain Monte Carlo (MCMC) modeling, yielding uncertainties capped at 0.9~km~s$^{-1}$ at most for CARMA~7 (as detailed below).

\section{Results and Discussion}
\label{Sec.Results}

We report the detection of the $J_{-1}\rightarrow(J-$ 1)$_{0}-E$ methanol lines at $J=7$ and $J=10$ in the following objects: CARMA-7, L1641N, NGC\,2024, IRAS\,16293, Serpens\,FIRS, and Serpens\,SMM4. A summary of these sources' properties can be found in Appendix\,\ref{sec:sources} while the mentioned methanol line transitions are shown in Fig.\,\ref{Fig.maser_specs}. In the following, we will refer to this sub-sample.

We observed a wide range of methanol transitions and specifically chose unblended emissions for inclusion in Table \ref{tab.methanol_lines}, where line parameters were taken from the Cologne Database for Molecular Spectroscopy \citep[CDMS;][]{Mueller2005}\footnote{\url{https://cdms.astro.uni-koeln.de/}}. After confirming the detection of these selected transitions, our next task was to investigate whether they adhere to local thermodynamic equilibrium (LTE) conditions. As a first approach, we can use the rotation diagram method and then perform radiative transfer modeling in LTE using the best fit of our rotation diagrams as an initial condition. If the transition moves beyond LTE, it will exceed both the rotation diagram best fit and the synthetic modeling within LTE. In such instances, a subsequent attempt involves replicating maser line emission by considering synthetic spectra beyond LTE and examining for negative optical depths. If negative optical depths are present, the line is inverted; if not, and the line remains unreproduced within LTE, the transition is in quasi-thermal emission.

\subsection{Rotation diagrams}

The rotation diagram method involves some assumptions, including the premise that all emissions uniformly fill the entire beam (refer to Appendix~\ref{sec.rot_diags_detailed} for further details). We emulate a filling factor of unity by setting a source size of 500\arcsec\, for the creation of the LTE models presented in Fig.~\ref{Fig.maser_specs} (dashed blue lines). This simplification is necessary due to the unknown extent of molecular gas emission, particularly from methanol tracing dense gas and weak shocks \citep[][and references therein]{Humire2022}. Figure \ref{fig.carma7_map} shows a significant aperture difference between our lines at $\sim$181-190 and $\sim$338~GHz for one of our sources as an example case. Despite this, our rotation diagrams perform well once transitions outside LTE (quasi-thermal and maser emission) are identified and discarded from the fit. Concerning quasi-thermal emission lines, exhibiting non-LTE behavior but with positive optical depths, our radiative transfer model, discussed in the following section, illustrates the essential role of a non-LTE component in reproducing their emissions. This is particularly evident in Figure~\ref{fig.nonLTECARMA7}, showcasing lines at 338.344 and 338.408~GHz where non-LTE emissions are predominant (indicated by the green lines in the figure). We therefore identify these couple of lines as quasi-thermal.

Our rotation diagrams are plotted in Fig.\,\ref{Fig.1}, where maser lines are in red, quasi-thermal lines in blue, and LTE lines in green. In the ISM, deviations from LTE are likely to occur in molecules with complex level diagrams, particularly when radiative and collisional processes vie for dominance in excitation and de-excitation. However, given the success in the fitting (see below), it can be inferred that potential blending lines in Table\,\ref{tab.methanol_lines} are, in most cases, not significantly contaminating. The fitting results, including 1$\sigma$ uncertainties, are labeled inside each subplot of Fig.\,\ref{Fig.1}.

Including a couple of highly improbable blended transitions with upper-level energies ($E_{\rm{up}}$) exceeding 240\,K, Table\,\ref{tab.methanol_lines} showcases the unblended methanol transitions detected in our sub-sample. We obtained rotation diagrams using the transitions presented in the mentioned Table from inside the CASSIS\footnote{\url{http://cassis.irap.omp.eu}} software. For this purpose, we have used the CDMS catalog, which does not discriminate between methanol A and E forms. This is because we do not have enough transitions for such a separation to be possible in creating rotation diagrams. 

Since blended transitions above an energy upper level ($E_{\rm up}/k$) of 150~K tend to be dominated by their lower-energy counterparts, as we thoughtfully inspected in a previous study \citep{Humire2022}, for the creation of the rotation diagrams we concentrate on methanol transitions with $E_{\rm up}/k<$150\,K. Moreover, we do not consider methanol lines separated from each other by less than their full width at half maximum (FWHM): when there is a scarce blending likelihood, we discard the transition from the global fit.

From the analysis presented above, we report the detection of the $7_{-1}\rightarrow 6_{0}-E$ and $10_{-1}\rightarrow 9_{0}-E$ methanol transitions as maser emission for the first time. To be rated as masers, the lines shall be out of LTE by more than 3$\sigma$ from our rotation diagram's best fits (and also have negative optical depths, see next Section). The rotation diagram's 1$\sigma$ is the $\rm{ln}($$N_{\rm{up}}/G_{\rm{up}})$ dispersion of our LTE transitions (green points in Fig.~\ref{Fig.1}), which accounts for the difference between our best fit and the observed values. Maser transitions in both lines have been detected in CARMA-7 and L1641N (see Fig.\,\ref{Fig.1}). For NGC\,2024 and Serpens FIRS, only the $10_{-1}\rightarrow 9_{0}-E$ transition exceeds LTE conditions without any doubt. Both in NGC\,2024 and Serpens FIRS the $10_{-1}\rightarrow 9_{0}-E$ line profiles are not as clear as for CARMA-7 and L1641N, having a low signal-to-noise ratio for NGC\,2024 and a large line width in Serpens FIRS, compared to other methanol transitions. 

Although we would expect to measure both the $7_{-1}\rightarrow 6_{0}-E$ and $10_{-1}\rightarrow 9_{0}-E$ maser lines with the same departure from LTE conditions, this is not the case in our sample. In CARMA-7 and L1641N, the $J=$7 transition departs more strongly from LTE conditions than the $J=$10 one. This can be explained in terms of the differences in frequency, with the $J=$7 line having $\sim$1.8 times smaller frequency than the $J=$10 line, implying $\sim$6 times more time to accumulate the inverted population (see also Sect.~\ref{sec.intro}). On the other hand, in NGC~2024 and Serpens~FIRS, the $J=7$ line lies even below quasi-thermal emission (blue points in Fig.\,\ref{Fig.1}), and only the $10_{-1}\rightarrow 9_{0}-E$ transition line can be considered as maser in these sources. One possible explanation for the latter could be the higher ($\times$1.8, in linear scale) angular resolution we use to observe the $J=$10 line. 

Finally, we also have the case of Serpens SMM~4, where only the $J=$7 transition exceeds the 3$\sigma$ fit. However, depending on the continuum level we choose for the Gaussian fitting in creating the rotation diagrams, this transition may or may not be considerably (3$\sigma$) beyond LTE, in addition, this transition does not surpass the two quasi-thermal lines (denoted by blue points in the rotation diagrams). We thus only determine a tentative detection for the $7_{-1}\rightarrow6_{0}-E$ line in Serpens SMM~4.

Among the mentioned sources, CARMA-7 exhibits the most prominent inverted population detection in the $7_{-1}\rightarrow 6_{0}-E$ transition, considering a 3$\sigma$ separation. Even when quasi-thermal emission is included in the LTE fit, along with error bars and all detected methanol transitions, as depicted in its rotation diagram (middle upper panel in Fig.\,\ref{Fig.1}), the mentioned $J=$7 transition in CARMA-7 remains beyond LTE conditions by more than 3$\sigma$. In these rotation diagrams, the departure from LTE (observed upper level column density over the expected one from the LTE best fit) for maser emission in CARMA-7 is 1.25 dex. At the same time, for L1641N, it is 0.83 dex. These values reflect the amplification effect due to negative optical depths, rather than indicating the actual upper level column density of the source.

\subsection{LTE plus non-LTE modeling}

Since CARMA-7 is the source where the $7_{-1}\rightarrow 6_{0}-E$ and $10_{-1}\rightarrow9_{0}-E$ methanol transitions show the clearest departures from LTE conditions, we focus on its spectrum to create radiative transfer models. Using RADEX \citep{vanderTak2007} (non-LTE) or LTE modeling separately does not reproduce our observations. However, using both models together, specifying the interacting mode on CASSIS, we can adjust the spectra, with the sum of those models reaching the expected values. A detailed explanation of our model procedure can be found in Appendix\,\ref{sec.details_model}, while this section focuses on its results.

Similar to previous work on methanol masers \citep{Humire2022}, we find that E-CH$_3$OH is more abundant than its A-CH$_3$OH counterpart for the LTE components. This is striking because the E-to-A methanol column density ratio (hereafter ISO, as it is called in CASSIS for "isotopic ratio"\footnote{\url{https://cassis.irap.omp.eu/docs/CassisScriptingDoc/setup/components_creation.html}}) should be less than or equal to unity at production temperatures up to approximately 40~K, and exactly unity at temperatures above this threshold \citep{Wirstrom2010}. Forcing the code to take an ISO of up to 1.0 for all the components never replicates several E-CH$_3$OH lines in the 7$_k\rightarrow$6$_k-E$ methanol series, such as those at $\sim$338\,GHz. 

\begin{table*}[!t]
\caption{Best-fit parameters from our LTE plus non-LTE modeling for CARMA-7.}
\label{tab.apen.sophisticated_models_params}
\begin{center}
\begin{tabular}{cccccccc}

\hline \hline
Component & $N$(Sp)              & $T_{\rm kin}$ or $T_{\rm ex}$  & FWHM         & $V_{\rm LSR}$   & $n_{\rm H_2}$& ISO & Size\\
&   \multicolumn{1}{c}{[$\times$10$^{13}$cm$^{-2}$]}              & \multicolumn{1}{c}{[K]} & \multicolumn{1}{c}{[km\,s$^{-1}$]} & \multicolumn{1}{c}{[km\,s$^{-1}$]}  & \multicolumn{1}{c}{[cm$^{-3}$]} & & \multicolumn{1}{c}{[$\arcsec$]}\\ \hline \\
non-LTE         & 41.49$^{+13.87}_{-2.48}$ & 99.60$^{+4.36}_{-0.64}$          &3.86$^{+0.18}_{-0.03}$          &0.90$^{+0.01}_{-0.01}$      & 7.27$\times$10$^{4}$$^{+0.23}_{-0.22}$ & 0.95$^{+0.06}_{-0.02}$ & 6.28$^{+0.32}_{-1.18}$\\
LTE (narrow)    & 79.48$^{+8.77}_{-6.62}$ & 15.85$^{+1.13}_{-0.21}$        &3.36$^{+0.04}_{-0.01}$          &0.73$^{+0.01}_{-0.03}$ & --        & 1.33$^{+0.13}_{-0.02}$& 9.21$^{+0.18}_{-0.68}$ \\
LTE (broad)     & 5.7$^{+0.10}_{-0.33}$  & 40.48$^{+0.40}_{-6.55}$        &13.21$^{+0.84}_{-0.11}$           &0.84$^{+0.01}_{-0.01}$& --        & 1.35$^{+0.04}_{-0.02}$& 41.44$^{+0.95}_{-1.21}$ \\
\hline \\
\end{tabular}
\tablefoot{Uncertainties correspond to 3$\sigma$. Velocity uncertainties are of the order of the instrumental uncertainty ($\sim$1.0~km\,s$^{-1}$). Column density ($N$(Sp)) values for A-type methanol can be obtained by dividing the E-type methanol values shown in the table by the ISO number given at the next-to-last column, while the rest of the parameters are the same for both methanol symmetric types. The dash symbol (--) indicates that no $n_{\rm H_2}$ was derived as the model is in LTE. An LSR velocity of 7$\pm$1.0~km~s$^{-1}$ (see Sect\,\ref{Sec.data_reduction} and Fig.\,\ref{Fig.maser_specs}) was previously subtracted. Therefore, the real $V_{\rm LSR}$ value, of about 7.8$\pm$1.0~km~s$^{-1}$, depending on the component, is obtained by summing up these values.}
\end{center}
\end{table*}

\begin{figure*}[!ht]
\centering
\includegraphics[width=0.9\textwidth, trim={3.8cm 24cm 6cm 2cm}, clip]{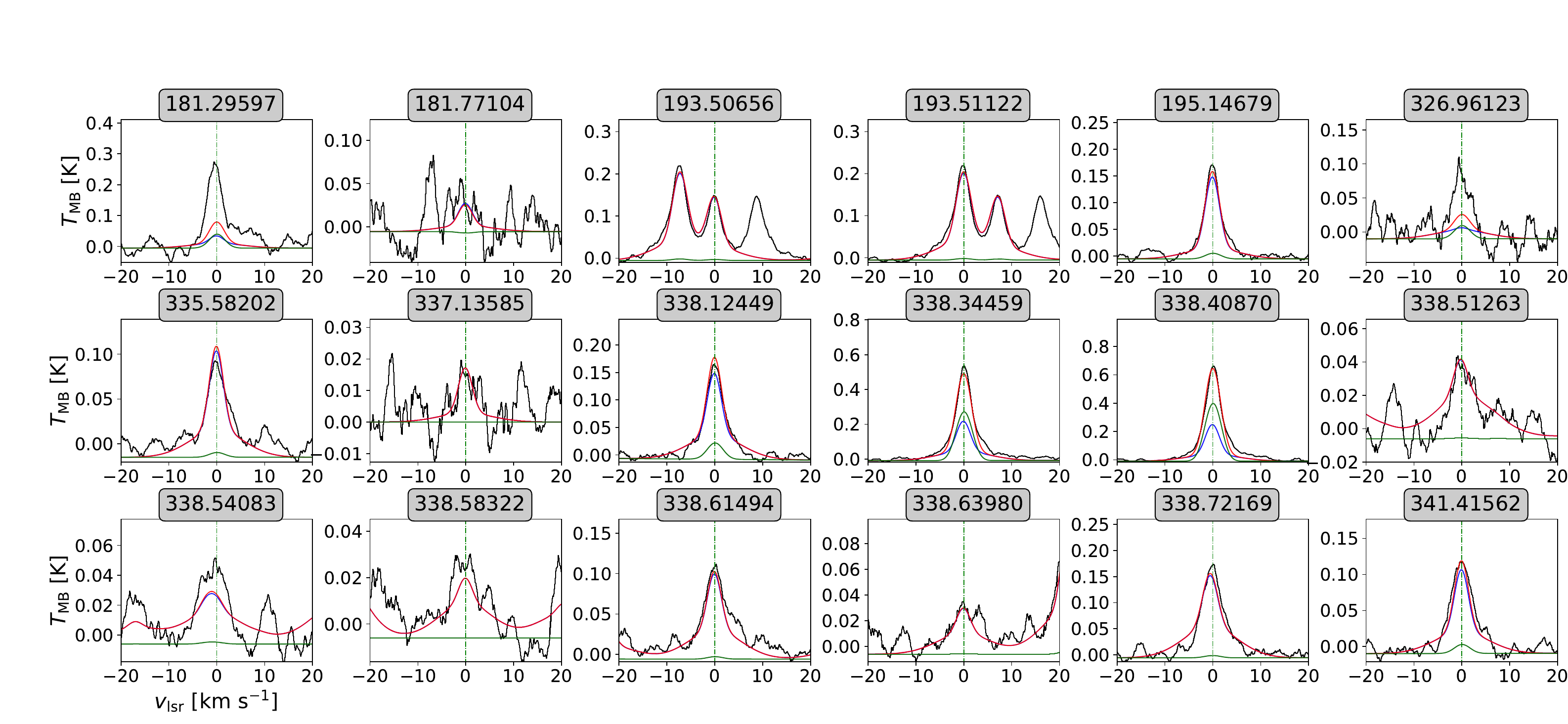}
\caption{Non-LTE plus LTE modeling for the CARMA-7 spectrum obtained from the best-fitting results (see Table~\ref{tab.apen.sophisticated_models_params}). Line rest frequencies are indicated at the top of each sub-panel. Rest-frame velocities and temperatures are labeled at the bottom and left-side positions in each sub-panel. Observed spectra are shown in black, LTE modeling (narrow plus broad components) is indicated in blue, RADEX modeling is in green, and the combination of LTE plus RADEX emission is in red. An LSR velocity of $\sim$7.8$\pm$1.0~km~s$^{-1}$ (see Sect\,\ref{Sec.data_reduction}, Fig.\,\ref{Fig.maser_specs}, and Table~\ref{tab.apen.sophisticated_models_params} for the used references, initially adopted and calculated $V_{\rm{LSR}}$, respectively) was previously subtracted.}
\label{fig.nonLTECARMA7}
\end{figure*}

A possible cause of ISO values above unity may come from the shocked nature of the methanol emission we observe, producing a mixture of methanol molecules from different environments along our line of sight. The latter is also a good argument for the line width and temperature of the broad LTE component. On the other hand, the narrow LTE component is the densest and coldest counterpart, which may arise from the central continuum source. This leads us to believe that the methanol emission we observe comes from a mixture of stellar outflows (broad LTE component and maser emission) and the central continuum source (narrow LTE component) in CARMA-7 (see also Appendix\,\ref{sec.details_model}), depicted in Fig.\,\ref{fig.carma7_map}.

In Fig.\,\ref{fig.nonLTECARMA7} we demonstrate how the non-LTE plus LTE approach reproduces the observations in CARMA-7. Table~\ref{tab.apen.sophisticated_models_params} presents the best-fit parameters. The $7_{-1}\rightarrow 6_{0}-E$ transition line falls approximately into the middle of the isotropic MMcIs luminosity distribution detected in low-mass star-forming regions \citep{Kalenskii2013}. Therefore, the newly detected masers likely represent a broadening in the maser luminosity vs. source luminosity relation of the so-far detected Class I maser population toward low-luminosity sources (see Sect.\,\ref{Sec.MMcIisolumvsbol}). 

As stated in the introduction, masers in high-$J$ transitions are hard to find due to the difficulty of keeping inverted populations. Among all the transitions detected by us in the spectrum of CARMA-7, we found only two transitions with negative optical depths, both in the $J_{-1}\rightarrow(J-$ 1)$_{0}-E$ family series: the $7_{-1}\rightarrow 6_{0}-E$ line at 181.295\,GHz and the $10_{-1}\rightarrow 9_{0}-E$ at 326.961\,GHz. Due to our coarse resolution and software limitations (see \citealt{vanderTak2007}, their Sect.\,3.6), we cannot ensure precise excitation temperatures and optical depths in our non-LTE models. Instead, we can assess the need for negative optical depths regarding the proposed maser emission. For the case of CARMA-7, the $10_{-1}\rightarrow 9_{0}-E$ line is quite beyond LTE conditions. At the frequencies of the lines observed with SEPIA345 ($\geq$323.153\,GHz) the beam size is 19.2". Therefore, we can conclude that the emission comes predominantly from CARMA-7 and not from the neighboring source CARMA-6, whose nucleus can not be completely disentangled in the low-frequency regime observed with SEPIA180 (indicated by an orange circle in Fig.\,\ref{fig.carma7_map}). Based on the CASSIS outputs for our non-LTE model, we derive excitation temperatures (and optical depths) of $-5.85$ ($-0.12$) and $-12.10$ ($-0.009$) in the $J=7$ and $J=10$ lines, respectively.

The resulting parameters for the three components of the best-fitting model are presented in Table~\ref{tab.apen.sophisticated_models_params}. From the inner to the outer layers, these components can be physically associated to the following: The warmest one comes from the shocked environment where Class~I maser spots emerge (see stellar outflows in Fig.\,\ref{fig.carma7_map}). The narrow LTE component is associated with the coldest and denser gas of the main source, corresponding to the ALMA 1~mm continuum source presented in Fig.~\ref{fig.carma7_map}. The broad LTE component is linked to an external warm and more diffuse (less dense) envelope around CARMA~7. This last component is affected by the shocked environment present in CARMA-7, which can account for its larger line widths, temperature, and source extension. The resulting synthetic spectrum is shown in Figure~\ref{fig.nonLTECARMA7}, where we separate between LTE (narrow plus broad, in blue lines) and non-LTE (green lines) emissions. More details about the model and results can be found in Appendix~\ref{sec.details_model}.

The evolution of the different free parameters vs. iteration number in our models, considering all components, is shown in Fig.~\ref{fig.parameters_vs_model_number}. Corner plots of our non-LTE and LTE narrow and broad components modeling are shown in Figs.\,\ref{fig.CARMA7_cornerplots_non-LTE}, \ref{fig.CARMA7_cornerplots_LTE}, and \ref{fig.CARMA7_cornerplots_LTE_broad}, respectively. 


\subsection{Maser luminosity vs. bolometric luminosity}
\label{Sec.MMcIisolumvsbol}
As pointed out by the review of \citet{Kalenskii2013}, the isotropic luminosity of Class I methanol masers at 44\,GHz is proportional to the bolometric luminosity of the object harboring it. This relation holds for low- to intermediate- and high-mass protostars. A long-monitoring study on low-mass protostars  \citep{Kalenskii2017} indicates that maser intensity in these objects does not change significantly over time, with small line profile changes rather attributed to calibration uncertainties than to intrinsic properties of the sources. 

The above-mentioned proportionality is plotted in our Figure\,\ref{fig.maser_int_vs_bol_luminosity}, where we conservatively associate 10\% variability to the masers, given the $\geq$20\% short-term (hours to days) variability found in 44~GHz MMcIs arising from massive star-forming regions \citep{Pratap2007}. New data from this study is added to literature information of \citet{Bae2011} previously taken by \citet{Kalenskii2013}. As can be seen in Figure\,\ref{fig.maser_int_vs_bol_luminosity}, we find that the relation is not fulfilled for high-$J$ transitions found in this study, as the isotropic luminosities measured for L1641N and CARMA-7 are two orders of magnitude stronger than expected by the mentioned relation.

Accounting for $L_{\rm{bol}}$ uncertainties, we include an average luminosity dispersion encountered in the literature for L1641N. In the case of CARMA-7, the difference between its bolometric luminosity inferred from 70$\mu$m observations and the bolometric luminosity of the closest source around the source position is included (see Appendix\,\ref{sec:sources}). 
Since neither \citet{Bae2011} nor \citet{Kalenskii2013} provide $L_{\rm{bol}}$ uncertainties for the whole sample included in our Figure\,\ref{fig.maser_int_vs_bol_luminosity} (grey dots), we conservatively assume an $L_{\rm{bol}}$ uncertainty of 21\%. The latter is motivated by the recent study of \citet{Pitts2022} who found $L_{\rm{bol}}$ uncertainties in the range of 3\% to 35\% (their Table\,4), with a mean of 21\% for a sample of intermediate- to high-mass star-forming regions.

\section{Conclusions}
\label{Sec.Conclusions}

In the present article, we report the first detections of the methanol $7_{-1}\rightarrow 6_{0}-E$ and $10_{-1}\rightarrow 9_{0}-E$ transitions out of LTE, which we consider as maser emission. These masers have been detected without any doubt in the low-mass star-forming regions CARMA-7 and L1641N. Tentative detections of the $7_{-1}\rightarrow 6_{0}-E$ transition are additionally claimed in NGC\,2024, Serpens FIRS, and Serpens SMM~4. In NGC\,2024 and Serpens FIRS, the $10_{-1}\rightarrow 9_{0}-E$ transition line departs from LTE ($\gg3\sigma$) and is also interpreted as maser emission.

\begin{acknowledgements}
We thank the anonymous referee for their helpful comments, questions, and suggestions on revising the manuscript. P.K.H. gratefully acknowledges the Fundação de Amparo à Pesquisa do Estado de São Paulo (FAPESP) for the support grant 2023/14272-4 as well as the support from the Joint ALMA Observatory (JAO) visitor program for facilitating this research. P.K.H. acknowledges the help of E. Caux through the CASSIS modeling process. G.N.O.L. acknowledges financial support from UNAM-DGAPA postdoctoral fellowship program.
\end{acknowledgements}

\bibliographystyle{aa}
\bibliography{aanda} 

\appendix

\section{Individual sources}
\label{sec:sources}

The whole sample used in the present study consists of low-mass YSOs located in our Galaxy.

\begin{figure}
\centering
\includegraphics[width=0.45\textwidth, trim={0cm 0.1cm 0cm 0}, clip]{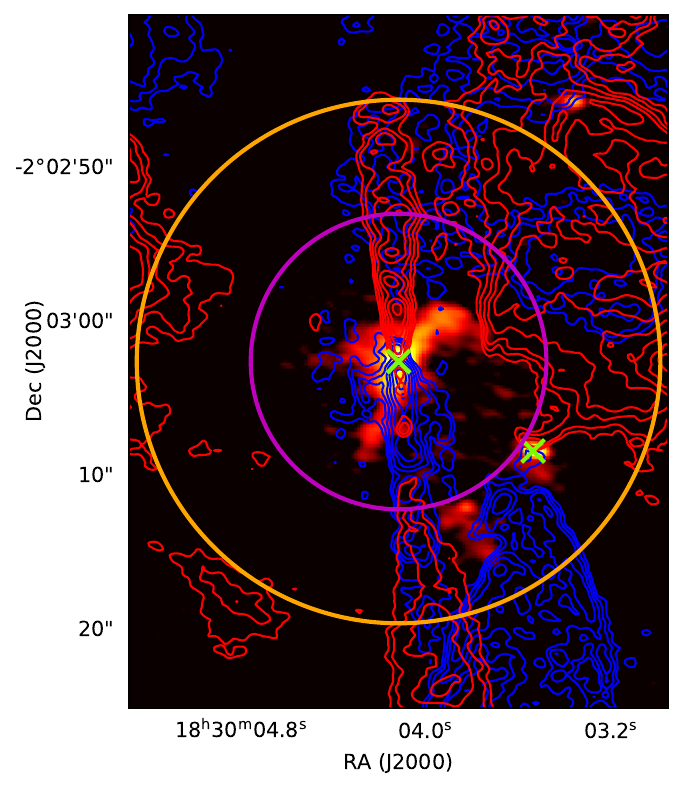}
\caption{CO ($J$=2--1) emission associated with the large-scale outflow of CARMA-7 (central red and blue contours; \citealt{Plunkett2015}). The integration ranges are -20 to 4\,km\,s$^{-1}$ for the blueshifted component and 12-40~km~s$^{-1}$ for the redshifted component. The $n$th contour occurs at a level of ($\sqrt{2}$)$^{n} \times p$ times the maximum signal strength, where the maximum signal strengths are 3.5 and 6.3 Jy~beam$^{-1}$~km~s$^{-1}$ and $p$ represents 10\% and 9\% of the signal strength for the blueshifted and redshifted emissions, respectively. Here, $n$ ranges from 0 onwards, incrementing by integers. The background is an ALMA map of 1\,mm continuum emission \citep{Plunkett2018}. APEX has a half-power beam width of 34\arcsec\ at 181.310 GHz (denoted by an orange circle) and 19.2\arcsec\ at 325.153 GHz (denoted by a magenta circle). The companion source whose center is also indicated by a green cross in the southeast is CARMA-6.}
\label{fig.carma7_map}
\end{figure}

\textbf{CARMA-7} is the strongest radio source of the Serpens South protostellar cluster. It presents a bipolar outflow which extends $\sim$0.16\,pc north-south (PA$\sim$4$^{\circ}$ east of north) in CO emission (Fig. \ref{fig.carma7_map}; \citealt{Plunkett2015}) and also shows water maser emission at 22\,GHz \citep{Ortiz-Leon2021}. We can adopt a distance of 440.7$\pm$4.6\,pc for CARMA-7, as recently measured by \citet{Ortiz-Leon2023} from H$_{2}$O maser emission at 22~GHz in CARMA-6.

Considering the closest Class 0 object to CARMA-7, identified as SerpS-MM18 by \citet{Maury2011} at an angular resolution of 11\arcsec, if we assume it is CARMA-7, it would likely have a bolometric luminosity ($L_{\mathrm{bol}}$) of 46$L_{\odot}$ at a distance of 440.7~pc. This estimation is however much larger than the one obtained for CARMA-7 with the Herschel 70\,$\mu$m band (8\arcsec) measurements, conducted by \cite{Podio2021}. Upon adjusting for the updated distance measured by \cite{Ortiz-Leon2023}, this latter $L_{\mathrm{bol}}$ becomes 20\,$L_{\odot}$. Hence, we may assume a $L_{\mathrm{bol}}$ range of 20 to 46\,$L_{\odot}$ for CARMA-7. Given that \cite{Podio2021} presented the highest angular resolution, we will consider their value as the more accurate and the one of \citet{Maury2011} as the upper limit.

\textbf{L1641N(orth)} is a dark cloud located in the southern part of our nearest giant molecular cloud, the Orion A region, located $\sim$7.2~pc southward from the Orion nebula cluster, assuming a distance of 414~pc \citep{Grossschedl2018}. Studies of FUV irradiation in protoplanetary disks indicate that L1641N presents the oldest population of stars in the L1641 region \citep{Terwisga2023}. 

Based on $^{12}$CO $J=$1--0 observations, \citet{Nakamura2012} inferred cloud-cloud collisions and protocluster winds. The latter may explain the presence of CO shells roughly centered at L1641N. The authors attribute the existence of multiple shells either to a fluctuating star formation rate or a single shell expanding into an inhomogeneous medium.

Following \citet{Fischer2017}, who detected 173 protostars in L1641, the median $L_{\rm{bol}}$ in L1641 is 2.0$\pm$4.7$L_{\odot}$. We will adopt this value for L1641N since the luminosity distribution in the entire complex is quite homogeneous among object classes.\\

\textbf{NGC2024} is a nearby HII region located in Orion B, and constitutes its most active and youngest star-forming region, with a median age of 0.5\,Myr. \citep{Levine2006}. VLBA measurements determine distances ranging between $\sim$350 and 540\,pc, depending on the binary system \citep{Kounkel2017}.

Among the several far infrared cores (FIR) located along its north-south distribution, we observed the brightest and oldest FIR\,5 source (VLA\,10 in \citealt{Rodriguez2003}, and references therein). FIR\,5 is the driving source of a unipolar molecular CO jet which collimates as its velocity increases up to 45\,km\,s$^{-1}$ (reaching an opening angle as low as $\sim$2$^{\circ}$; \citealt{Richer1992}). 

Judging for common bolometric luminosities measured around this source by \citet{Haisch2001}, we can assume values in the 2--3\,$L_{\odot}$ range, where the closest source, numbered 58 by the mentioned study, shows a $L_{\rm{bol}}$ of 2.3$L_{\odot}$. Chandra observations did not detect any emission source, likely because it is heavily obscured by dust \citep{Skinner2003}.\\

\textbf{Serpens~SMM4} is a Class\,0 protostellar system with an inverse P-Cygni profile initially claimed to be an infall signature \citep{Narayanan2002} but lately associated with large-scale cloud flows or foreground emission \citep{Mottram2013}. Being part of the Serpens South cluster, its distance can be approximated to 440.7$\pm$4.6\,pc \citep{Ortiz-Leon2023}.\\

\textbf{Serpens FIRS 1} is a Class\,0 protostar located in the main core of the Serpens Molecular Cloud at a distance of 436$\pm$9\,pc \citep{OrtizLeon2018}. Also known as Serpens SMM\,1, this protostar is associated with a bipolar radio jet \citep[e.g.,][]{Curiel1996}. It is the most embedded, massive, and luminous YSO in the Serpens dark cloud, with a total flux of 5.9\,Jy at 1.1\,mm, a mass of 15\,M$_{\odot}$ \citep{Enoch2007} and a bolometric luminosity of 91\,$L_{\odot}$ \citep{Bae2011}, after correcting for the updated distance. High angular resolution observations at 0\farcs6 revealed a second YSO indicating a binary configuration for this system \citep{Choi2012}.\\

\textbf{IRAS~16293--2422} is a well-studied Class 0 hierarchical system of solar type protostars, composed by two main condensations, IRAS~16293--2422A and IRAS~16293--2422B, surrounded by an extended envelope of $\sim$ 8000 AU \citep{Crimier2010, Jacobsen2018}. It is located in the L1689N region of the $\rho$ Ophiuchus cloud at a distance of 141~pc \citep{Dzib2018}. This multiple system shows a wealth of molecular species at both small \citep{Jorgensen2016} and large scales \citep{Kahle2023}. In addition, a multiple outflow system has been observed toward this source \citep{Mizuno1990,Girart2014}. 


\section{Rotation diagrams}
\label{sec.rot_diags_detailed}

\begin{figure*}[!ht]
\centering
\includegraphics[width=0.99\textwidth, trim={0 0 0 0}, clip]{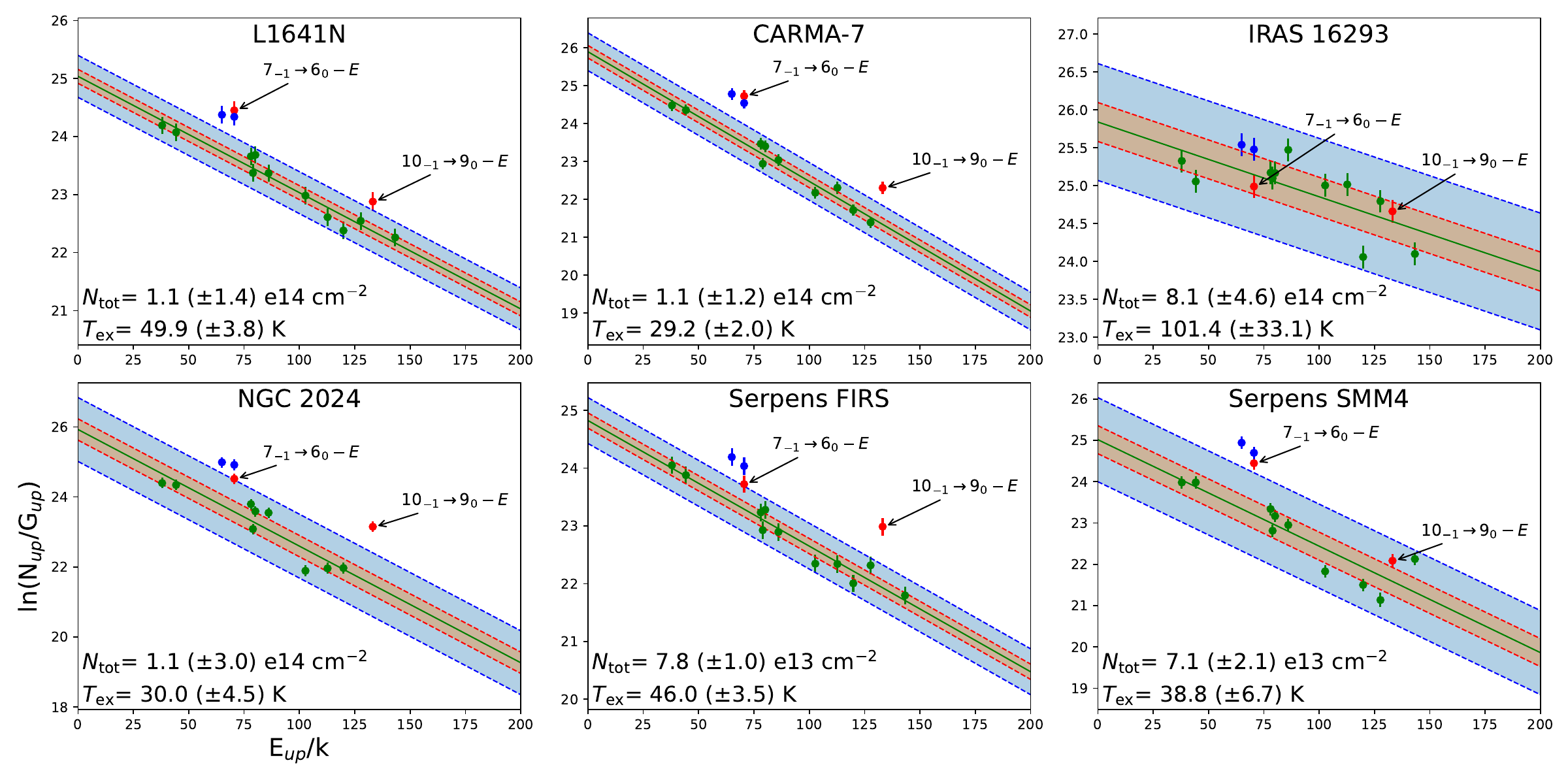}
\caption{Rotation diagrams for sources with detected $7_{-1}\rightarrow 6_{0}-E$ methanol transition (indicated with an arrow). The shaded red and blue regions represent the uncertainty level at 1 and 3$\sigma$, respectively. Methanol lines in the $J_{-1}\rightarrow(J-$ 1)$_{0}-E$ line series, prone to be masers, are in red, thermal lines are in green, and quasi-thermal emission (those out of LTE best fit, positive optical depths and strong presence of a non-LTE component in our models) are in blue. These quasi-thermal transitions correspond to the 7$_{-1}\rightarrow$6$_{-1}-E$ and 7$_{0}\rightarrow$6$_{0}-A^{+}$ transition lines at 338.344588 and 338.408698\,GHz, respectively. Only unblended lines were used for the fit. Best fit results (total column densities ($N_{\rm{tot}}$) and excitation temperatures ($T_{\rm{ex}}$) plus 1$\sigma$ uncertainties) are labeled in the bottom left corner of each subplot. Transitions detected in each source are listed in Table\,\ref{tab.methanol_lines}.}
\label{Fig.1}
\end{figure*}

We use rotation diagrams as our first and simplest method to unveil the presence of methanol masers. Its construction assumes LTE conditions, a filling factor of unity, and that the lines are optically thin. It also assumes the Rayleigh-Jeans (RJ) approximation\footnote{see, for instance, the derivation presented in \citet{Araya2005}, between Equations A9 and A10}, valid when $\nu~[\rm{GHz}] \ll 20.84~\it{T}_{\rm{ex}} [\rm{K}]$ \citep[e.g.,][]{Wilson2009}. Given that the lowest excitation temperature derived by this method is 29.2~K (see CARMA-7 in Fig.~\ref{Fig.1}), the RJ approximation is valid for frequencies much lower than $\sim$610~GHz, which is the case of this study. The rotation diagram method also assumes a negligible background continuum \citep[e.g.,][]{Belloche2019}. Since we subtracted the continuum emission before our analysis (see Sect.~\ref{Sec.data_reduction}), this assumption does not represent a problem for us. More details about this method can be found in \citet{Goldsmith1999}.

\begin{table*}[!thb]
\caption{Selected methanol transitions.} 
\label{tab.methanol_lines}
\scriptsize
\renewcommand{\tabcolsep}{0.16cm}
\begin{center}
\begin{tabular}{lllllllllll}
\hline \hline
Transition                  & $\nu$ [GHz]& A/E       & $E_{\rm{up}}/k$ [K]&L1641N&CARMA-7 & IRAS\,16293& NGC\,2024  & Serpens FIRS  & Serpens SMM4 & potential blending lines\\ \hline \\
7$_{-1}\rightarrow$6$_{0}$  & 181.295971 & E         & 70.55     & \checkmark & \checkmark & \checkmark & \checkmark & \checkmark    & \checkmark   & $^{13}$CC$_{2}$N, CH$_3$OH with $E_{\rm{up}}$=1025.4\,K\\
4$_{1}\rightarrow$3$_{1}$ & 193.506559 & E         & 44.26     & \checkmark & \checkmark & \checkmark & \checkmark & \checkmark    & \checkmark   & \\
4$_{1}\rightarrow$3$_{1}$   & 195.146790 & A$^{-}$   & 37.95     & \checkmark & \checkmark & \checkmark & \checkmark & \checkmark    & \checkmark   & CH$_{3}$OCHO\\ 
9$_{1}\rightarrow$9$_{0}$   & 322.239462 & A$^{-+}$  & 119.88    & \checkmark & \checkmark & \checkmark & \checkmark & \checkmark    & \checkmark   & -- \\
10$_{1}\rightarrow$10$_{0}$ & 326.630583 & A$^{-+}$  & 143.27    & --         & --         & \checkmark & --         & \checkmark    & \checkmark   & -- \\
10$_{-1}\rightarrow$9$_{0}$ & 326.961232 & E         & 133.15    & \checkmark & \checkmark & \checkmark & \checkmark & \checkmark    & \checkmark   & S$^{17}$O, H2CCHCN\\
7$_{1}\rightarrow$6$_{1}$   & 335.582017 & A$^{+}$   & 78.97     & \checkmark & \checkmark & \checkmark & \checkmark & \checkmark    & \checkmark   & -- \\
7$_{0}\rightarrow$6$_{0}$ & 338.124488 & E         & 78.08     & \checkmark & \checkmark & \checkmark & \checkmark & \checkmark    & \checkmark   & -- \\
7$_{-1}\rightarrow$6$_{-1}$   & 338.344588 & E         & 70.55     & \checkmark & \checkmark & \checkmark & \checkmark & \checkmark    & \checkmark   & -- \\
7$_{0}\rightarrow$6$_{0}$   & 338.408698 & A$^{+}$   & 64.98     & \checkmark & \checkmark & \checkmark & \checkmark & \checkmark    & \checkmark   & CH$_3$OH with $E_{\rm{up}}$=243.8\,K \\
7$_{-3}\rightarrow$6$_{-3}$   & 338.559963 & E         & 127.71    & \checkmark & \checkmark & \checkmark & --         & \checkmark    & \checkmark   & -- \\
7$_{3}\rightarrow$6$_{3}$ & 338.583216 & E         & 112.71    & \checkmark & --         & \checkmark & \checkmark & \checkmark    & --           & $^{33}$SO, CH$_{3}$SH\\
7$_{1}\rightarrow$6$_{1}$ & 338.614936 & E         & 86.05     & \checkmark & \checkmark & \checkmark & \checkmark & \checkmark    & \checkmark   & CH$_{3}$OCHO\\
7$_{2}\rightarrow$6$_{2}$   & 338.639802 & A$^{+}$   & 102.72    & \checkmark & --         & \checkmark & \checkmark & \checkmark    & \checkmark   & C$_{2}$H$_{5}$C--13--N\\
7$_{1}\rightarrow$6$_{1}$   & 341.415615 & A$^{-}$   & 80.09     & \checkmark & \checkmark & \checkmark & \checkmark & \checkmark    & \checkmark   & cis--DCOOH, n--C$_{3}$H$_{7}$CN\\ \hline  
\end{tabular}
\tablefoot{Selected methanol lines used to perform our rotation diagrams (Fig.\,\ref{fig.carma7_map}). Possible contamination with other molecules is mentioned in the rightmost column. $\checkmark$ indicates well-distinguished/confidently-detected lines while the dash symbol (--) indicates a poor signal or non-detection. Certain methanol species that are highly improbable to blend with others are also identified: The 7$_{-1}\rightarrow$6$_{0}-E$ line falls within the range of frequencies of the 19$_{2}\rightarrow$18$_{3}-E$ methanol line in all sources while the 7$_{0}\rightarrow$6$_{0}-A^{+}$ methanol transition falls within the frequency range of the 7$_{6}\rightarrow$6$_{6}-E$ methanol transition for the case of IRAS\,16293 and CARMA~7, given the higher FWHM of the CH$_{3}$OH lines in these sources compared to the rest of the sample. Line frequencies and upper-level energies were taken from the CDMS database \citep{Mueller2005}.}
\end{center}
\end{table*}

\section{Details about the model}
\label{sec.details_model}

Our procedure to create the CASSIS modeling was as follows: We started with a single LTE model with the rotation diagram best fits as inputs, and then we added a non-LTE component attempting to reproduce the quasi-thermal and maser emission. We selected a slab geometry for this latter as this configuration is the most appropriate for shocks \citep[see, e.g.,][]{Leurini2016}. The quasi-thermal transitions correspond to the 7$_{-1}\rightarrow$6$_{-1}-E$ and 7$_{0}\rightarrow$6$_{0}-A^{+}$ transition lines at 338.344588 and 338.408698\,GHz, respectively. They depart ($>$3$\sigma$) from LTE conditions in our rotation diagrams for the case of L1641N, CARMA~7, NGC~2024, Serpens FIRS, and Serpens~SMM~4, as seen in Fig.\,\ref{Fig.1}, and could not be fitted by our LTE model alone. Therefore, we included them for the LTE plus non-LTE modeling as inputs.

Since masers are non-LTE phenomena, they were also not reproduced by our single LTE modeling. This corresponds to the $7_{-1}\rightarrow6_{0}-E$ and $10_{-1}\rightarrow9_{0}-E$ methanol transition lines at 181.295 and 326.961~GHz, respectively. We did not try to fit these lines in our LTE plus non-LTE modeling since maser emission can depart far from the non-LTE conditions obtained using RADEX. On the other hand, we reproduced the quasi-thermal transitions with a mixture of LTE plus non-LTE components (see Fig.~\ref{fig.nonLTECARMA7}).

Both the $7_{-1}\rightarrow6_{0}-E$ and $10_{-1}\rightarrow9_{0}-E$ transitions were not reproduced by our LTE plus non-LTE models, as seen in Fig.~\ref{fig.nonLTECARMA7}. Even so, we note that the $J=$7 transition seems to be more diluted in its corresponding APEX beam (34\arcsec), as depicted in Fig.~\ref{fig.carma7_map}, than its higher-$J$ transition counterpart, the $J=$10 line, observed with a beam of $\sim$19.2\arcsec, and which presents a more peaked profile. The same effect can be inferred from the line profiles in L1641N and NGC~2024 in Fig.~\ref{Fig.maser_specs}.

Initially, we fixed a $V_{\rm{LSR}}$ of 7.82~km~s$^{-1}$ for the LTE component, derived from previous attempts with all parameters being free. The ISO value was set to vary freely since the lowest reduced $\chi^{2}$ ($\chi^{2}_{red}$) were only obtained with an ISO greater than unity. Testing with several independent models, we found that an ISO greater than unity is particularly critical to reproduce the $7_{k}\rightarrow6_{k}-E$ methanol transitions with $k= 0$ and $+$1 at 338.124488 and 338.344588~GHz, respectively. After finding good solutions for the LTE model, we added a non-LTE (RADEX) component to it aiming to reproduce quasi-thermal emission. We reached convergence after 1000 iterations using a constant $V_{\rm{LSR}}$ of 7.8755~km~s$^{-1}$, which was obtained from previous modeling after reaching convergence. This way, we obtained an acceptance rate of 0.42. The same LTE plus non-LTE model delivers an ISO of 1.11, that slightly varied across the 1000 iterations (between 1.02 and 1.12). We fixed the $V_{\rm{LSR}}$ to 7.88 and the ISO value to the expected 1.0 upper limit \citep{Wirstrom2010} for the non-LTE modeling.

Aiming at reproducing the broad component in our spectra, most clearly seen in the 4$_{k}\rightarrow$3$_{k}$~$A^{+}/E$ transitions (with $k=$1 and $-$2), at $\sim$193.51~GHz (top middle panels of Fig.\,\ref{fig.nonLTECARMA7}), we added an extra broad LTE component with an FWHM of 13.9~km~s$^{-1}$, an ISO of 1.4, and a $V_{\rm{LSR}}$ of 7.83~km~s$^{-1}$. To obtain these initial conditions we freely vary the broad LTE component after fixing the previously encountered best values for the LTE plus non-LTE models over the parameter space.

Finally, after several attempts including interacting and non-interacting conditions, we obtained the best initial conditions presented in Table~\ref{tab.apen.table_initial_conditions}, allowing all of them to vary freely within the indicated values.

\begin{table*}[!t]
\caption{Initial conditions used for our CARMA-7 model.}
\label{tab.apen.table_initial_conditions}
\begin{center}
\begin{tabular}{cccccccc}

\hline \hline
Component & $N$(Sp)              & $T_{\rm kin}$ or $T_{\rm ex}$  & FWHM         & $V_{\rm LSR}$   & $n_{\rm H_2}$& ISO & Size\\
&   \multicolumn{1}{c}{[$\times$10$^{13}$cm$^{-2}$]}              & \multicolumn{1}{c}{[K]} & \multicolumn{1}{c}{[km\,s$^{-1}$]} & \multicolumn{1}{c}{[km\,s$^{-1}$]}  & \multicolumn{1}{c}{[cm$^{-3}$]} & \multicolumn{1}{c}{[$\times$10$^{4}$cm$^{-2}$]} & \multicolumn{1}{c}{[$\arcsec$]}\\ \hline \\
non-LTE         & 51.35 (10--200) & 102.96 (40--140) &3.83 (2.5--4.9) &0.90 (0.5--0.95) & 8 (4--20) & 1.0 (0.6--1.5) & 4.62 (3--30)  \\
LTE (narrow)    & 83.97 (7-500)    & 17.74 (15--60)    &3.27  (3--43)   &0.71 (0.6--0.88) & --        & 1.6 (0.6--3.0) & 8.92 (3--20)   \\
LTE (broad)     & 7.24 (2--100)   & 25.15 (15--60)    &13.89 (10--23)  &0.83 (0.75--0.9) & --        & 1.4 (0.6--1.5) & 41.8 (20--70)  \\
\hline \\
\end{tabular}
\tablefoot{Initial conditions and range of permitted values provided parenthetically for the parameters in our LTE plus non-LTE radiative transfer modeling of CARMA-7.}
\end{center}
\end{table*}


Based on the resulting synthetic spectrum, we found the best solutions in the interacting mode. In this setup, CASSIS uses the spectrum calculated for the first component as the continuum background for the next one (E. Caux, priv. comm.).

Typically, denser and warmer regions are situated behind colder ones. Consequently, we designated the warmest non-LTE component as the initial layer. An exception to this arrangement occurs when the outer, more widespread foreground component experiences a temperature increase due to shocks. Considering that scenario, we assigned the narrow LTE component as the second layer, despite its best-fitting values indicating lower temperatures than those of the broad LTE component. This decision is substantiated by its more concentrated emission, as depicted in the ALMA continuum image of CARMA-7 (Fig.~\ref{fig.carma7_map}).

The broad LTE component was then positioned as the third and final layer. These components, arranged from the innermost to the outermost layers, can be associated with the following physical phenomena.

The warmest one can be associated with the shocked environment where Class~I maser emission emerges (see stellar outflows in Fig.\,\ref{fig.carma7_map}). As maser coherent emission mainly arises from the plane of the sky, this component possesses low line widths. The narrow LTE component is associated with the coldest and denser gas of the main source, showing the lowest line width as it is not affected by shocks. The broad LTE component is linked to an external warm envelope such as the ISM around CARMA~7 which, to some extent, may share CARMA~6 emission in the lowest frequency regime. Being so extended, its emission surpasses the APEX beam with a size of 41\farcs4, at these dimensions, we expect a significant contribution from the LTE components of the shocked environment present in CARMA-7, which can also account for the line widths of 13.2~km~s$^{-1}$. 

In the MCMC algorithm, the step size determines how far the walkers, which are a set of values for the variables (i.e., column densities, temperatures, sizes, etc.), move in a single iteration of the MCMC algorithm. A large step size allows a wider range of parameter values to be explored in each step but can lead to higher rejection rates; on the other hand, a smaller step size limits the movement of the walkers, something that can lead to more accurate estimation of the posterior distribution but slow the exploration of the parameter space and increase computational cost. Similar to the procedure described in \citet{Hunter2014}, we adjusted the step size based on the resulting acceptance rate\footnote{The acceptance rate in the context of MCMC sampling refers to the proportion of proposed moves that are accepted during the sampling process. It is calculated as the ratio of the number of accepted moves to the total number of proposed moves.} and established a cutoff parameter, set at half the total iterations, to control the step size.

\begin{figure}[!ht]
\centering
\includegraphics[width=0.49\textwidth, trim={0cm .3cm 0cm 0.9cm}, clip]{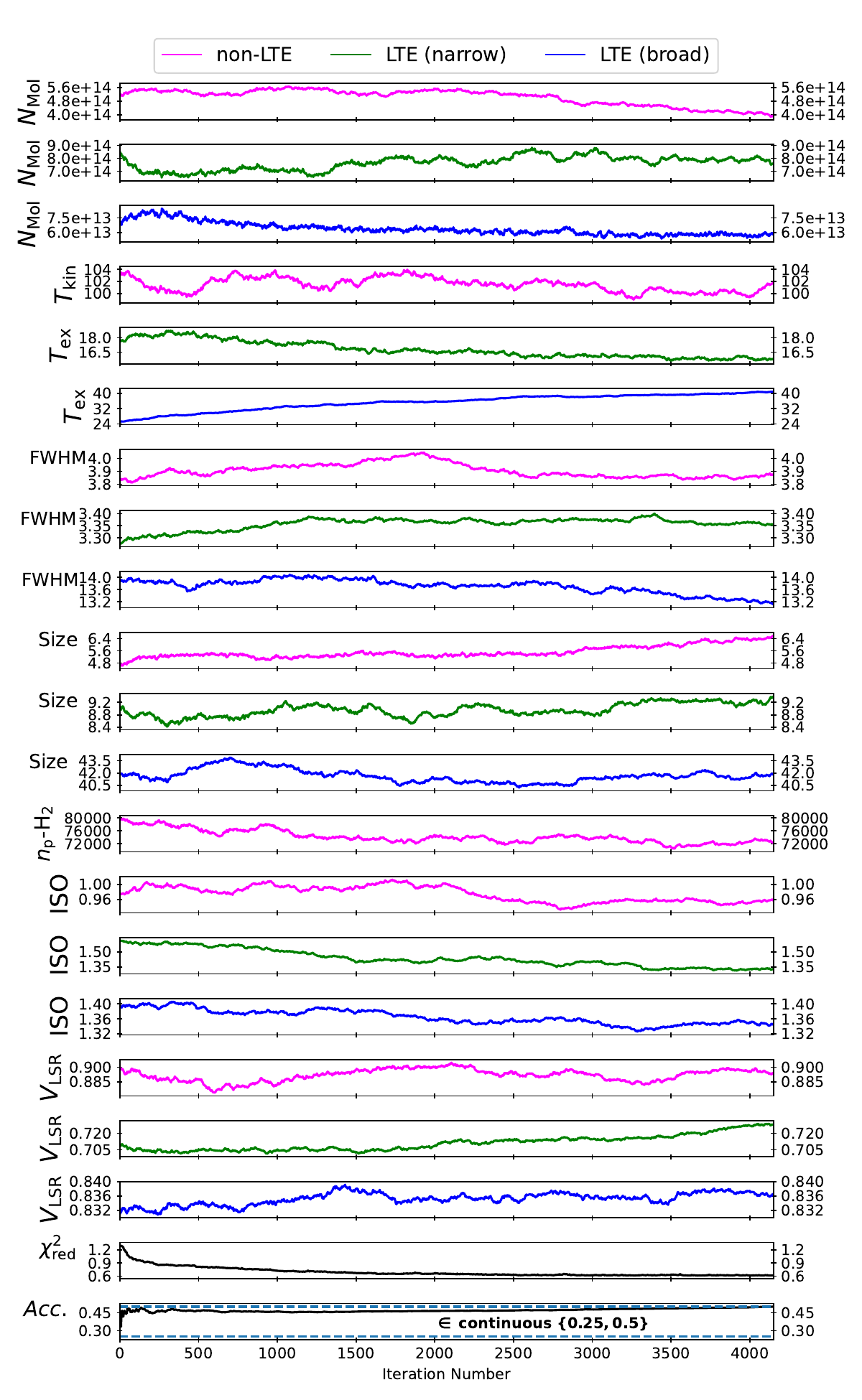}
\caption{Evolution of the free parameters in each component vs. model iteration number. 
}
\label{fig.parameters_vs_model_number}
\end{figure}

\begin{figure}[!ht]
\centering
\includegraphics[width=0.5\textwidth, trim={0cm 0cm 0cm 0}, clip]{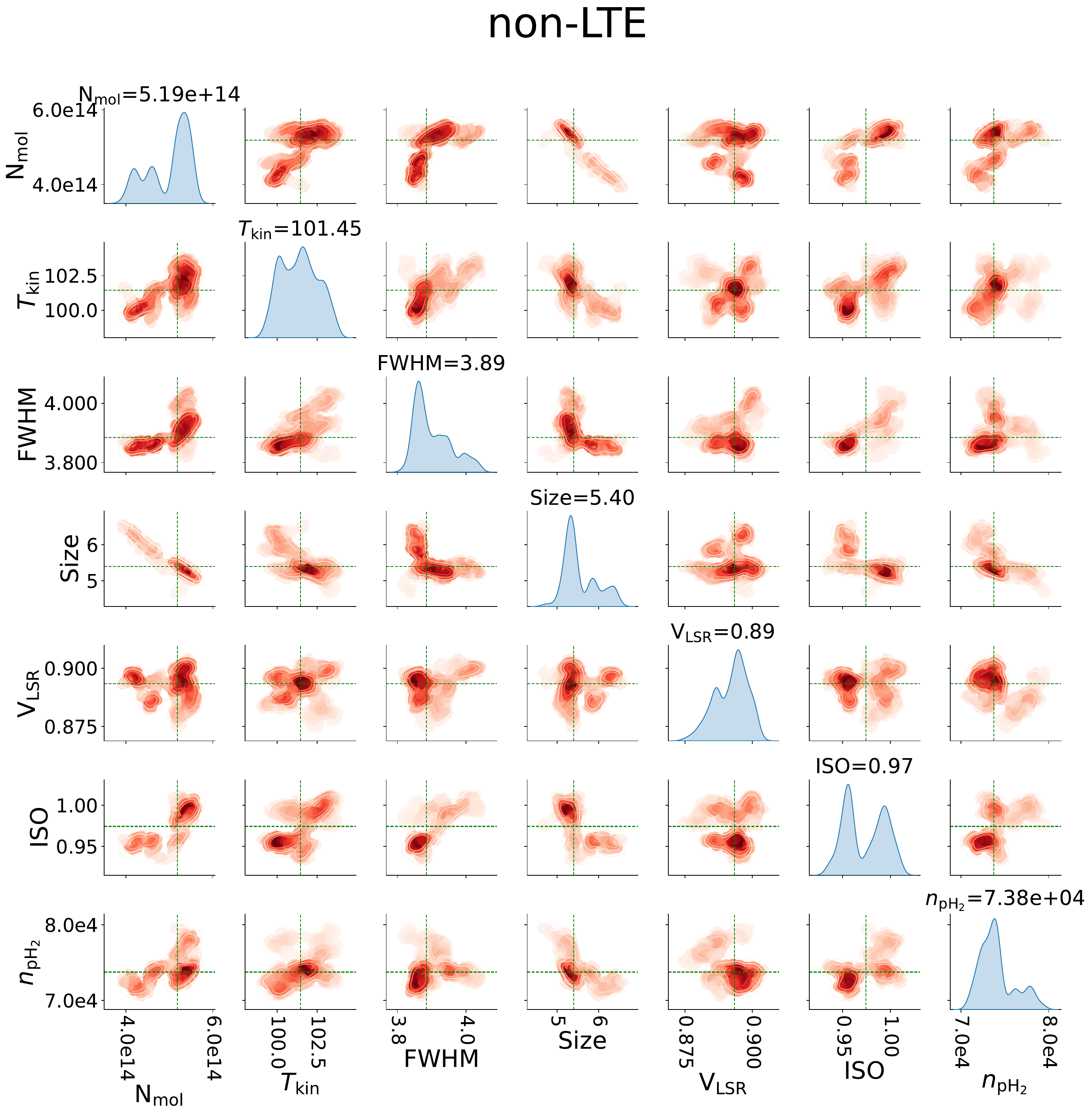}
\caption{Corner plots of the derived free MCMC parameters after 4151 iterations, only contiguous iterations with an acceptance rate in the 0.25--0.5 range are shown. This model was used to create the non-LTE best fit for CARMA-7 inside CASSIS. Median values are labeled above each Gaussian distribution and also marked by green dashed lines inside each subplot.}
\label{fig.CARMA7_cornerplots_non-LTE}
\end{figure}

\begin{figure}[!ht]
\centering
\includegraphics[width=0.5\textwidth, trim={0cm 0cm 0cm 0}, clip]{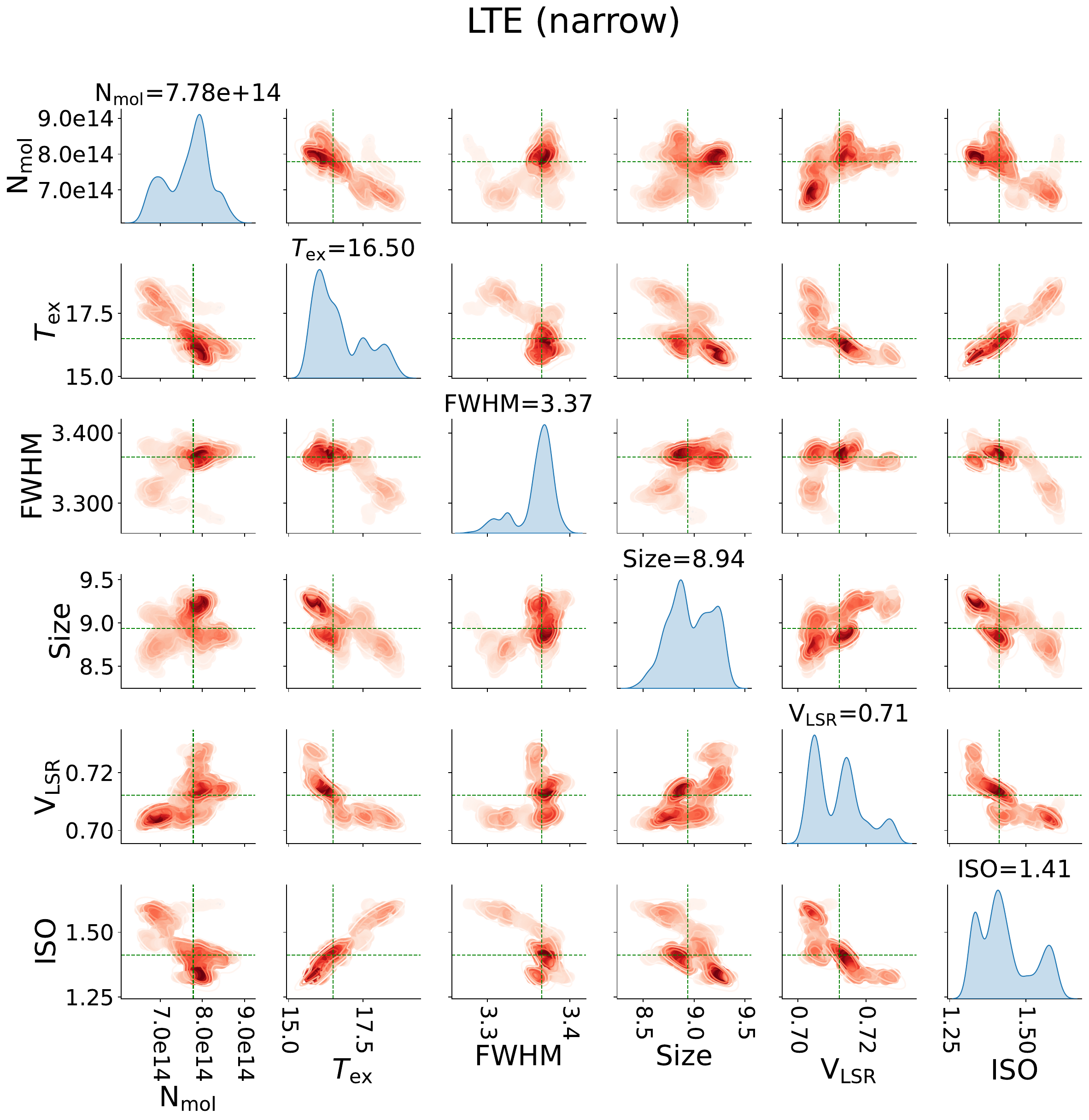}
\caption{Corner plots of the derived free MCMC parameters used to create the LTE narrow component best fit for CARMA-7 inside CASSIS, same as for Fig.\,\ref{fig.CARMA7_cornerplots_non-LTE} but with excitation temperatures instead of kinetic temperatures and devoid of molecular para-hydrogen density ($n_{\rm{p-H_{2}}}$) (see Appendix~\ref{sec.details_model}).}
\label{fig.CARMA7_cornerplots_LTE}
\end{figure}

\begin{figure}[!ht]
\centering
\includegraphics[width=0.5\textwidth, trim={0cm 0cm 0cm 0}, clip]{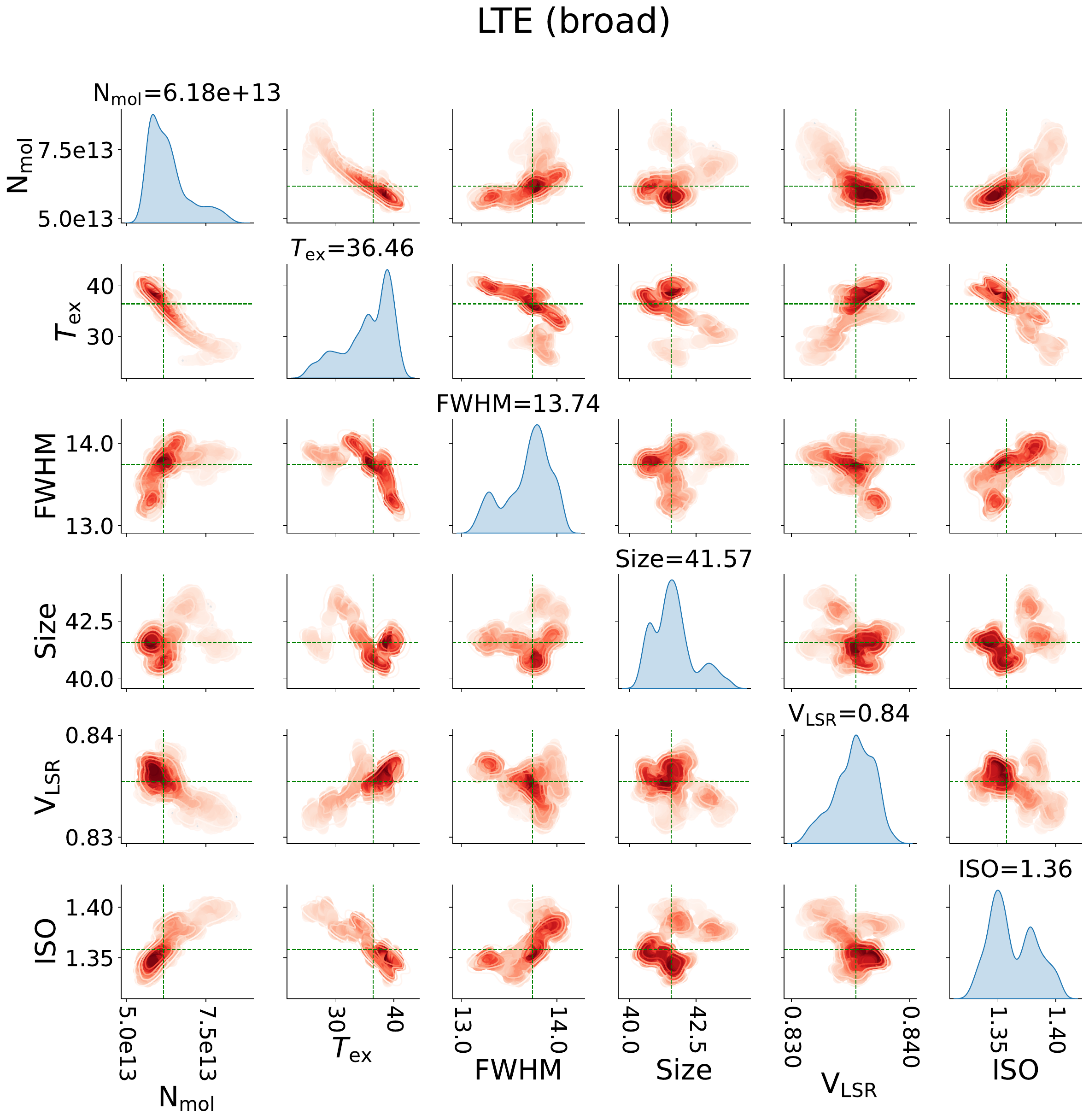}
\caption{Corner plots of the derived free Markov chain Monte Carlo (MCMC) parameters used to create the LTE broad component best fit for CARMA-7 inside CASSIS, same as for Fig.\,\ref{fig.CARMA7_cornerplots_LTE}.}
\label{fig.CARMA7_cornerplots_LTE_broad}
\end{figure}

We found good results after 4151 iterations with an acceptance rate of 0.49 using a step size of 200, and a $\chi^{2}_{red}$ of 0.61. The best-fit parameters are indicated in Table\,\ref{tab.apen.sophisticated_models_params}. In Fig.~\ref{fig.parameters_vs_model_number} we illustrate the evolution of the free parameters in our MCMC modeling throughout the model iteration number. We consider only those values where the acceptance rate (shown in the bottom panel of the Figure) consistently falls between 0.25 and 0.5 for the corner plots in Figures~\ref{fig.CARMA7_cornerplots_non-LTE} to \ref{fig.CARMA7_cornerplots_LTE_broad}. After 4151 iterations, our model continued till the maximum of 10000 iterations we initially set, but its acceptance rate increased over 0.5, reaching a final value of 0.57. These last 6000 iterations were therefore not included in the results.

\section{Comparison with previous work}
\label{sec.comparison}

To put our results in context, we consider the sample analyzed by \citet{Bae2011} regarding methanol masers at 44~GHz and integrate our new findings. While we explained the adopted bolometric luminosities in our Section \ref{Sec.MMcIisolumvsbol}, in the following we will provide details regarding the calculations of the isotropic maser intensities. 

Applying a single Gaussian fitting, we obtain peak temperatures of 0.2748 and 0.2474~K, and FWHMs of 3.7519 and 3.0833~km~s$^{-1}$ for the MMcIs$J_{-1}$ at $J=$7 in CARMA-7 and L1641N, respectively. With these values, we perform the following calculations to determine the maser isotropic luminosity:

First, we transform our peak temperature ($T_{\rm{peak}}$ in Kelvin) to Jy using the Kelvin to Jansky conversion factor described in the APEX website\footnote{\url{https://www.apex-telescope.org/telescope/efficiency/index.php.old}}. In our case, this factor equals 24.4/0.71, where 24.4 is the constant provided in the mentioned website and 0.71 is the aperture efficiency, $\eta_{a}$, for SEPIA180 (see Sect~\ref{Sec.data_reduction}); then we have:
\begin{equation}
    S_{\rm{peak}} = \frac{24.4}{0.71} T_{\rm{peak}},
\end{equation}

\noindent
where $S_{peak}$ is the peak flux density in Jy. After that, approximating the maser emission by a Gaussian profile, we can infer the integrated intensity using the FWHM:

\begin{equation}
    I = S_{\rm{peak}} V_{\rm{FWHM}}\sqrt{\frac{\pi}{4ln(2)}}.
\end{equation}

\noindent

To translate apparent luminosity from Jy~km~s$^{-1}$ to isotropic luminosity in $L_{\odot}$, we multiply by $10^{-23}$ for Janskys to erg~s$^{-1}$~cm$^{-2}$~Hz$^{-1}$ and by $10^{5}$ for km~s$^{-1}$ to cm~s$^{-1}$. Then we divide by the speed of light in cm~s$^{-1}$ and multiply by the line frequency in Hz. Finally, we divide by the solar luminosity (3.9$\times$10$^{33}$~erg~s$^{-1}$) and multiply by the square of the distance in cm to get the result in solar luminosities:

\begin{equation}
 L_{\rm{maser}} [L_{\odot}]= 4\pi d^{2} \nu I/c
\end{equation}
\noindent
Where the distance, $d$, is in cm and the integrated intensity $I$ is in $L_{\rm{\odot}}$~cm$^{-2}$ cm~s$^{-1}$. In the denominator, the frequency is in Hz, and $c$ is the speed of light in cm~s$^{-1}$. For CARMA-7 and L1641N we assume distances of 440.7~pc and 414~pc, respectively (see Appendix~\ref{sec:sources}). The above computes isotropic luminosities of 1.35$\times$10$^{-6}$ and 8.88$\times$10$^{-7}$ $L_{\odot}$ for CARMA-7 and L1641N, respectively.

\begin{figure}[!ht]
\centering
\includegraphics[width=0.49\textwidth, trim={0cm 0.2cm 0cm 0}, clip]{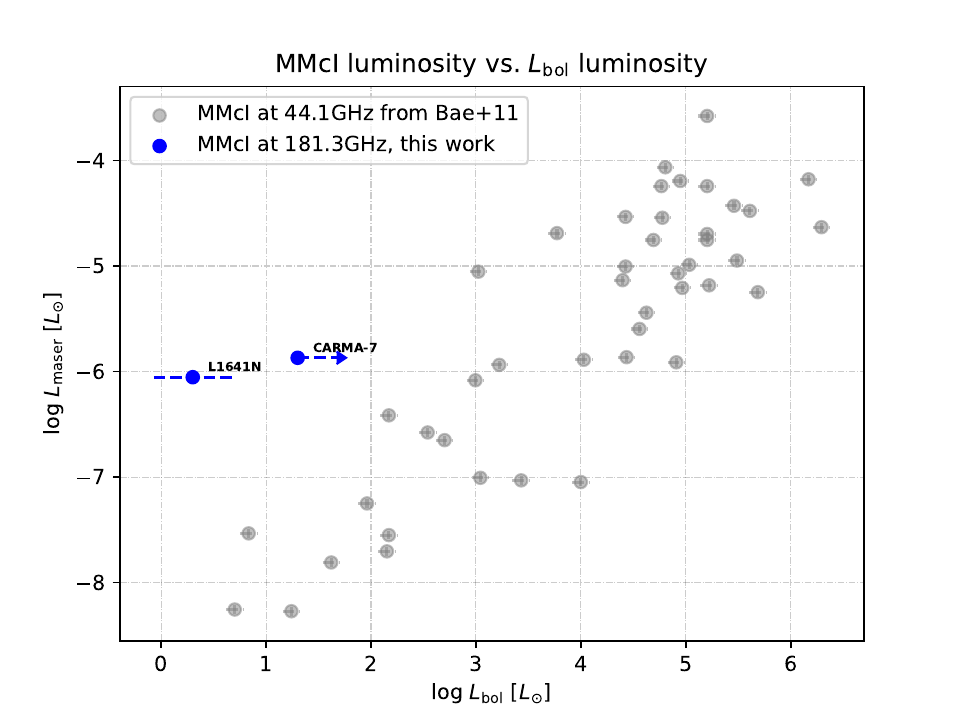}
\caption{Isotropic MMcIs luminosity at 181\,GHz in low-mass star-forming regions (blue points, this work) and archive data (grey points) on 44\,GHz MMcIs considering low- to high-mass star-forming regions, all this compared to the bolometric luminosity $L_{\rm{bol}}$ of the mentioned objects. Archive information comes from \citet{Bae2011}. These data were directly extracted from \citet[][their Fig.\,2]{Kalenskii2013}.}
\label{fig.maser_int_vs_bol_luminosity}
\end{figure}

\end{document}